\documentclass[conference]{IEEEtran}
\IEEEoverridecommandlockouts
\usepackage{cite}
\usepackage{amsmath,amssymb,amsfonts}
\usepackage{algorithmic}
\usepackage{graphicx}
\usepackage{textcomp}
\usepackage{xcolor}
\usepackage{colortbl}

\usepackage[a4paper, total={184mm,239mm}]{geometry}
\def\BibTeX{{\rm B\kern-.05em{\sc i\kern-.025em b}\kern-.08em
    T\kern-.1667em\lower.7ex\hbox{E}\kern-.125emX}}
\begin{document}

\title{Exploiting Kernel Compression on BNNs
}

\author{\IEEEauthorblockN{Franyell Silfa}
\IEEEauthorblockA{\textit{Department of Computer Architecture} \\
\textit{Universitat Politècnica de Catalunya}\\
Barcelona, Spain \\
fsilfa@ac.upc.edu}
\and
\IEEEauthorblockN{Jose Maria Arnau}
\IEEEauthorblockA{\textit{Department of Computer Architecture} \\
\textit{Universitat Politècnica de Catalunya}\\
Barcelona, Spain \\
jarnau@ac.upc.edu}
\and
\IEEEauthorblockN{Antonio González}
\IEEEauthorblockA{\textit{Department of Computer Architecture} \\
\textit{Universitat Politècnica de Catalunya}\\
Barcelona, Spain \\
antonio@ac.upc.edu}

}

\maketitle

\begin{abstract}

Binary Neural Networks (BNNs) are showing tremendous success on realistic image classification tasks. Notably, their accuracy is similar to the state-of-the-art accuracy obtained by full-precision models tailored to edge devices. In this regard, BNNs are very amenable to edge devices since they employ 1-bit to store the inputs and weights, and thus, their storage requirements are low. Moreover, BNNs computations are mainly done using xnor and pop-counts operations which are implemented very efficiently using simple hardware structures. Nonetheless, supporting BNNs efficiently on mobile CPUs is far from trivial since their benefits are hindered by frequent memory accesses to load weights and inputs.

 In BNNs, a weight or an input is stored using one bit, and aiming to increase storage and computation efficiency, several of them are packed together as a sequence of bits. In this work, we observe that the number of unique sequences representing a set of weights or inputs is typically low (i.e., 512). Also, we have seen that during the evaluation of a BNN layer, a small group of unique sequences is employed more frequently than others. Accordingly, we propose exploiting this observation by using Huffman Encoding to encode the bit sequences and then using an indirection table to decode them during the BNN evaluation. Also, we propose a clustering-based scheme to identify the most common sequences of bits and replace the less common ones with some similar common sequences. As a result, we decrease the storage requirements and memory accesses since the most common sequences are encoded with fewer bits.
 
 In this work, we extend a mobile CPU by adding a small hardware structure that can efficiently cache and decode the compressed sequence of bits. We evaluate our scheme using the ReAacNet model with the Imagenet dataset on an ARM CPU. Our experimental results show that our technique can reduce memory requirement by 1.32x and improve performance by 1.35x.

{\let\thefootnote\relax\footnote{\textit{{Accepted for publication in DATE 2023.}}}}

\end{abstract}


\section{Introduction}

Deep Neural Networks (DNNs) have recently achieved tremendous success in various tasks such as image classification and speech recognition. Not surprisingly, they have become the most critical component for Machine Learning Applications. 
Nonetheless, the cost of storing and evaluating DNNs is high since they usually are composed of millions of parameters. Mainly, evaluating DNNs on edge devices is quite challenging due to the constrained resources on edge devices. In this regard, Binary neural networks (BNNs) are an attractive solution to reduce the cost of evaluating DNNs on edge devices. In BNNs, the weights and input features can only take the value of $1$ and $-1$; thus, one bit is used to store them. Therefore, the cost of storing a BNN layer can be reduced dramatically compared to those that use integer or floating point values. Moreover, BNNs can be executed very efficiently with modest hardware since the matrix multiplications required to evaluate a DNN, can be performed using a series of xnor and popcount operations.

While BNNs are very computationally efficient, their accuracy tends to be relatively low compared to a similar state-of-the-art full-precision network. However, recent BNN models, such as ReactNet~\cite{reactnet} and FracBNN~\cite{zhang2021fracbnn}, achieve a top-1 accuracy comparable to state-of-the-art full-precision networks. For instance, ReActNet obtains a top-1 accuracy of 69\% on the Imagenet dataset while MobileNet~V2~\cite{mobilenetv2} (i.e., CNN model tailored to edge devices) has a top-1 accuracy of 72\%. Nevertheless, the number of bits required to store ReactNet is 29 million, whereas it is 27 million for MobileNet~V2. Hence, evaluating these models on edge devices is still challenging due to their limited storage and computational resources. Most of the storage and computational requirements of BNNs are due to convolutional operations.

In this work, we focus on improving the performance and storage requirement of BNN models using convolutional operations. In this regard, in a binary kernel (i.e., weights are 1 or -1), for any given \textit{n by n} channel stored as a sequence of bits, there are $2^{n*n}$ possible sequences of bits.
However, we observe that during the evaluation of a convolution operation for a given kernel, some sequences of bits are used more frequently than others. For instance, in some layers of ReActNet, 32 individual sequences of bits account for more than 50\% of all the possible sequences. To exploit this observation, we propose to employ a variable length compression algorithm to encode each sequence of bits. Then, we assign an encoding with fewer bits to the most common sequences since they are more likely to be used when evaluating a kernel. For simplicity in the rest of the paper, we refer to the sequence of bits formed with the values of a given channel as \textit{bit~sequence}.

 A significant challenge in implementing a variable-length encoding is providing a software- and hardware-friendly solution while also being efficient~\cite{phuffman}. A software-only implementation is insufficient since the performance is severely affected by the overhead of decoding the compressed bit sequences. We address this issue by providing hardware support to decompress the \textit{bit sequences}. Moreover, when evaluating BNNs on mobile CPUs, the matrix multiplications are done very fast since the latency of computing a xnor and a popcount is very low. Therefore, the loads to fetch the weights are in the critical path. Hence, we also provide hardware support to stream the weights automatically to mitigate this issue.

Another important observation is that for some layers, a less frequently used \textit{bit sequence} can be replaced by one that is employed more often without negatively impacting the accuracy of the network. Accordingly, we propose a clustering algorithm that substitutes some of the less common \textit{bit sequences} with others that are similar to them but also very common. By reducing the less common \textit{bit sequences}, we can compress the binary kernels more efficiently since the number of bits to store the most common ones is smaller.

In summary, the main focus of this paper is improving the performance and memory requirements of BNNs on mobile CPUs. We claim the following contributions:

\begin{itemize}
    \item 
We analyze the frequency distribution of all the \textit{bit sequences} in the network and show that some of them are used more frequently than the rest.

    \item  We propose to use a simple variable-length encoding to compress the binary kernels. Also, we describe a novel clustering algorithm that increases the frequency of the most common \textit{bit sequences}. Our scheme compresses the BNN layers by 1.32x on average and improves the model storage requirements by 1.2x.
    
    \item We propose simple hardware extensions and evaluate them on top of a modern ARM CPU. Our proposals improve the performance by 1.35x.

\end{itemize}

\section{Background}

\subsection{Binary Neural Networks}\label{s:bnn_networks}
Like full-precision DNNs, Binary Neural Networks (BNNs) are composed of many layers (i.e., convolution, dense, etc) stacked on top of each other. BNNs use one-bit weights and inputs constrained to $+1$ and $-1$, but they usually are trained using full-precision weights. Then, the weights and inputs are binarized using the following equation:      

\begin{equation}
x^b = \begin{cases}
+1    &\text{if }  x>=0, \\
-1         &\text{otherwise, }  
\end{cases}
\label{e:x_binarization}
\end{equation}

where $x$ is a full-precision weight or input and $x^b$ is the binarized value.

Convolution operations with binary inputs and weights are computed as their full-precision counterpart. However, to calculate an output feature Equation~\ref{e:binary_mac} is employed. In this equation, $o$ is a feature of the output feature map, whereas $w^b$ and $x^b$ are the binarized kernel (weights) and input vectors. Note that the kernel and the input are composed of $1$ and $-1$, but they are stored as $1$ and $0$, respectively.

\begin{equation}
o = popcount(xnor(w^b,x^b))
\label{e:binary_mac}
\end{equation}

Equation~\ref{e:binary_mac} is essentially computing a multiply-accumulate operation (MAC) which is the main computation on DNNs. Hence, as mentioned earlier, BNNs are very efficient in computations and memory requirements since MAC operations can be computed using xnor and popcount functions, and weights are stored using 1 bit. 

\subsection{BNN model}\label{s:cnn_model}
Recently, on image classification tasks, BNN models can achieve top-1 accuracy, similar to the accuracy of full-precision models tailored to edge devices. In this regard, ReActNet~\cite{reactnet} is a BNN model based on MobileNet~\cite{mobilenetv2} with a top-1 accuracy of 69.4\% on the ImageNet dataset. Note that the accuracy of Mobilenet V2 for the same task is 72\%. Since the accuracy of ReActNet is very close to the full-precision counterpart, we employ this model as the baseline in this work.    

\begin{figure}[t!]
    \centering
    \includegraphics[width=3.375in]{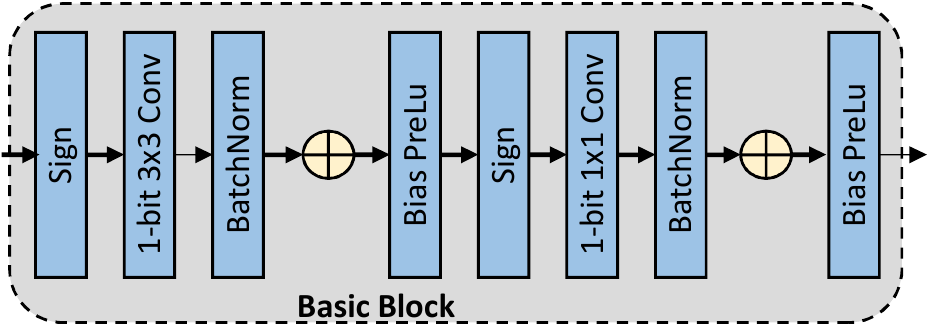}
    	\vspace*{-2mm}
    \caption{ReActNet basic block and Architecture.}
    \label{f:reacnet_arch}
\end{figure}

The main building block of ReActNet is shown in Figure~\ref{f:reacnet_arch}. This block comprises two convolutional operations that employ binarized input and kernels. Also, it performs other operations such as batch-norm and Prelu activation functions which are computed using full-precision. One of the critical contributions of this model is that the Prelu activation is biased by shifting and reshaping its input. By performing this transformation, the accuracy of the network is substantially improved~\cite{reactnet}.    

ReActNet is composed of 15 layers which include one input, one output layer, and 13 layers that are constructed using the basic block shown in Figure~\ref{f:reacnet_arch}. The input layer is convolutional, whereas the output layer is fully-connected. Both layers are computed using full-precision values, and in this work, we quantize them using 8 bits. 

Table~\ref{t:reacnet_storage} shows the storage requirements and execution time for the main operations in ReActNet. As can be seen, the storage requirement due to the 3x3 convolutions accounts for 68\% of the total. Furthermore, these 3x3 convolutions represent 67\% of the execution time; hence in this work, we focus on improving these convolutions.

 \begin{table}[t!]
	\caption{ReActNet storage and execution time breakdown. Others refers to activation and normalization layers,etc.}
	\label{t:reacnet_storage}
	\centering
	\begin{tabular}{cccc}
		\cellcolor[gray]{0.9}\small\textbf{Operation}&\cellcolor[gray]{0.9}\cellcolor[gray]{0.9}\small\textbf{Storage (\%)} &\cellcolor[gray]{0.9}\small\textbf{Precision (bits)}&\cellcolor[gray]{0.9}\small\textbf{Execution time(\%)}   \\
		\small Input Layer &\small 0.02&\small 8&\small 4.0     \\
		\cellcolor[gray]{0.9}\small Output Layer &\cellcolor[gray]{0.9}\small 22.17\ &\cellcolor[gray]{0.9}\small 8&\cellcolor[gray]{0.9}\small 18.7   \\
		\small Conv 1x1 &\small 8.5&\small 1&\small 6.9    \\
		\cellcolor[gray]{0.9}\small Conv 3x3 &\cellcolor[gray]{0.9}\small 68.0&\cellcolor[gray]{0.9}\small 1&\cellcolor[gray]{0.9}\small 66.8 \\
			\small Others &\small 1.31& \small 32& \small 3.6   \\	
		
	\end{tabular}
\end{table}

\section{Kernel Compression}

We observe that for the 3x3 binarized kernels found in ReActNet, each channel is composed of nine values that are either one or zero; thus,  there are only 512 unique $bit~sequences$. However, some $bit~sequences$ are rarely used, while other sequences tend to appear very frequently. Therefore, we could exploit this observation by employing a variable encoding that assigns fewer bits to the most common $bit~sequences$. Moreover, we experimentally found that some rarely used $bit~sequences$ can be substituted by some others that are more common without negatively affecting the overall network accuracy.

\begin{figure}[t!]
    \centering
    \includegraphics[width=3.375in]{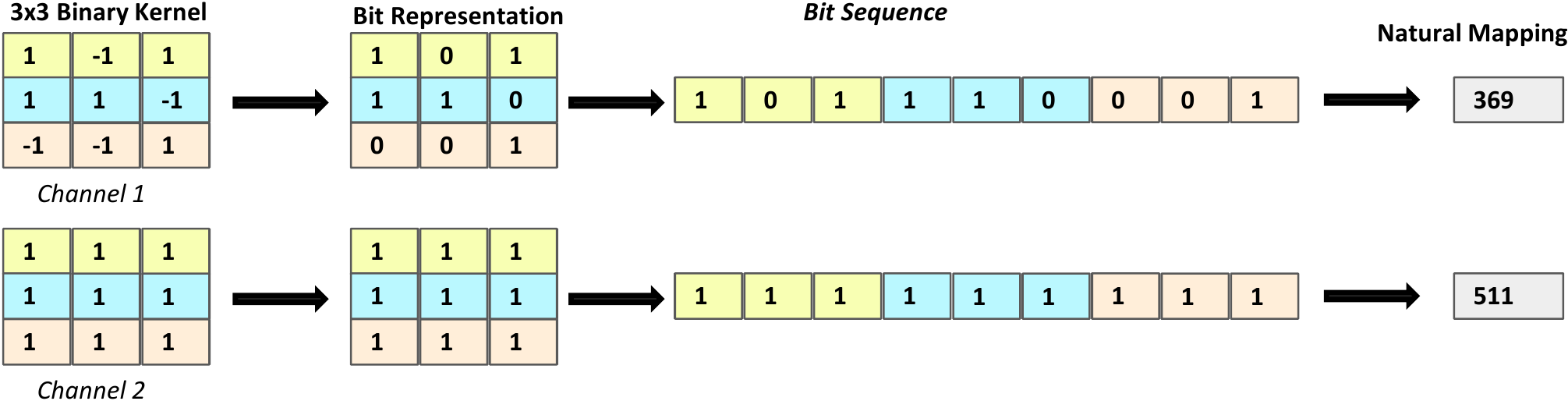}
    	\vspace*{-2mm}
    \caption{3x3 binary kernel with 2 channels. Conceptually the values of a channel are mapped to an integer number. }
    \label{f:unique_sequences_ex}
\end{figure}

Figure~\ref{f:unique_sequences_ex} shows a binarized 3x3 kernel with 2 channels. As mentioned earlier, each of these channels is composed of nine values which are either $1$ or $-1$, and to represent one channel a least 9 bits are needed. For clarity, to refer to a given $bit~sequence$,  we employ a Natural mapping, as shown in Figure~\ref{f:unique_sequences_ex}, that assigns a Natural number to each $bit~sequence$. In this regard, we assign the value in position 0,0 to the most significant bit and the value in position 2,2 to the least significant bit. For example, a channel where all the values are $zeros$ (i.e., -1) will be referred to as 0. Similarly, a channel where all the values are $ones$ will be mapped to 511. Note that this is a conceptual representation, and the actual layout of these bits in memory can be different.  

The rest of this section presents an analysis of the use frequency for each $bit~sequence$ in ReActNet. Then, we describe our methodology to replace uncommon $bit~sequences$. Finally, we detail our compressing scheme.

\subsection{Frequency of use for $bit~sequences$}\label{s:sequence_frequency}

\begin{figure}[t!]
    \centering
    \includegraphics[width=3.375in]{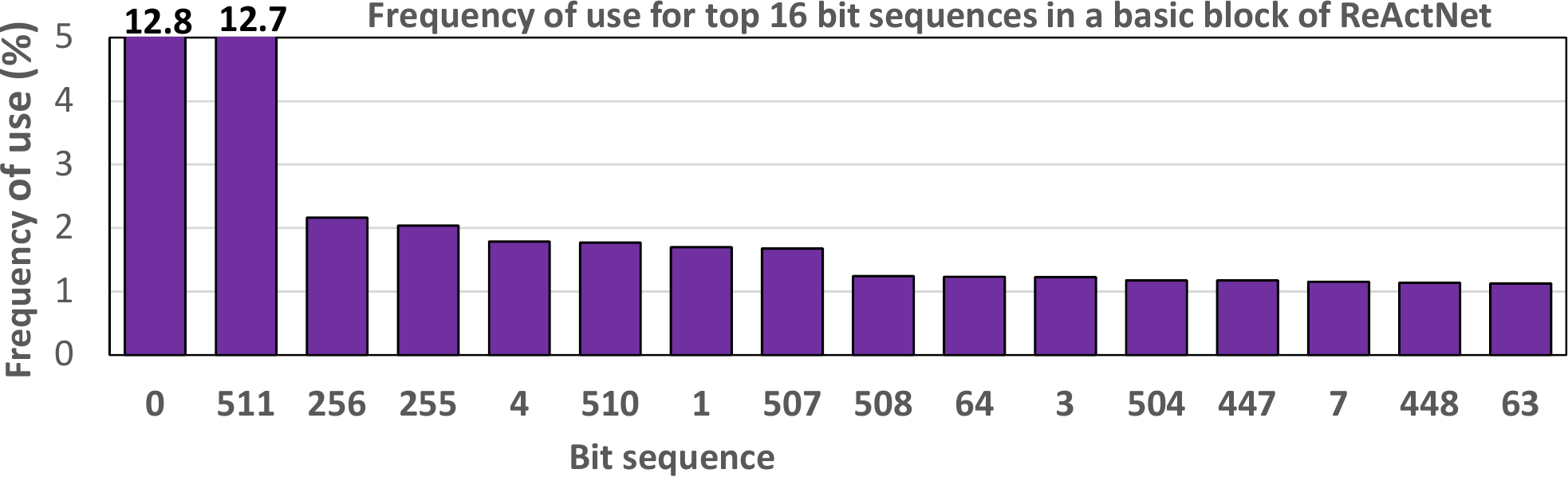}
    	\vspace*{-2mm}
    \caption{Frequency of use for the top 16 $bit~sequences$ for the 3x3 binary kernels in one of  ReActNet's basic blocks.}
    \label{f:frequency_sequence_l12}
\end{figure}

Figure~\ref{f:frequency_sequence_l12} shows the frequency of use for the 16 more common $bit~sequences$ in all the 3x3 binary kernels in one of the basic blocks of ReActNet.
As shown in the plot, $bit~sequences$ containing only zeros or ones account for 25\%. In contrast,  the most common 16 and 64 $bit~sequences$ account for 46\% and 75\% of all of them, respectively. Regarding the kernels in other basic blocks, their behavior is similar to the one in Figure~\ref{f:frequency_sequence_l12}. As shown in Table~\ref{t:distribution_by_layer}, in  all the blocks, the top 64 $bit~sequences$ account for more than
50\% of the sequences. Moreover, for all the blocks, the 256 most frequently used $bit~sequences$ account for up to 92\%. Henceforth, instead of employing nine bits to store each channel, we could use 5 bits to store half of them and nine or less for the rest.

 \begin{table}[t!]
	\caption{ Distribution of $bit~sequences$ for the 3x3 kernel in each basic block.}
	\label{t:distribution_by_layer}
	\centering
	\begin{tabular}{ccc}
		\cellcolor[gray]{0.9}\small\textbf{Layer}&\cellcolor[gray]{0.9}\small\textbf{Top 64 (\%)} &\cellcolor[gray]{0.9}\small\textbf{Top 256 (\%)}\\
		\small Block 1 &\small 53.4&\small 90.6\\
		\cellcolor[gray]{0.9}\small Block 2 &\cellcolor[gray]{0.9}\small 64.5 &\cellcolor[gray]{0.9}\small 95.1\\
		\small Block 3 &\small 56.3&\small~87.11\\
		\cellcolor[gray]{0.9}\small Block 4 &\cellcolor[gray]{0.9}\small 64.8&\cellcolor[gray]{0.9}\small 92.7\\
			\small Block 5 &\small 63.2& \small 88.3\\
		\cellcolor[gray]{0.9}\small Block 6 &\cellcolor[gray]{0.9}\small 63.1 &\cellcolor[gray]{0.9}\small 90.86 \\
		\small Block 7 &\small 62.4&\small 91.64\\
		\cellcolor[gray]{0.9}\small Block 8 &\cellcolor[gray]{0.9}\small 60.8&\cellcolor[gray]{0.9}\small 90.24\\
			\small Block 9 &\small 55.2& \small 92.9\\	
		\cellcolor[gray]{0.9}\small Block 10 &\cellcolor[gray]{0.9}\small 62.2 &\cellcolor[gray]{0.9}\small 89.9 \\
		\small Block 11  &\small 67.97&\small 92.\\
		\cellcolor[gray]{0.9}\small Block 12 &\cellcolor[gray]{0.9}\small 75.3&\cellcolor[gray]{0.9}\small 93.4\\
		\small Block 13  &\small 58.3&\small 86.9    
				
	\end{tabular}
\end{table}

\subsection{Compressing $bit~sequences$ }\label{s:kernel_compression}
Aiming to increase storage efficiency and to exploit the variability in the frequency of use of the $bit~sequences$, we propose to use Huffman Encoding to encode them. When using Huffman encoding to compress the kernels, we must first create a Huffman tree, where each $bit~sequence$ is assigned to a tree node. Then, before using a $bit~sequence$, we must decode it with the help of lookup tables that contain the uncompressed value. Note that the tree is created offline, whereas the decoding is done at runtime.      

\begin{figure}[t!]
    \centering
    \includegraphics[width=3.375in]{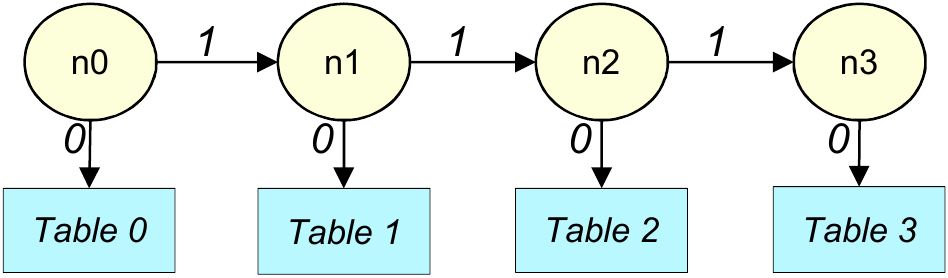}
    	\vspace*{-2mm}
    \caption{Simplified Huffman tree with four nodes. The prefix of an encoded bit sequence is used as a pointer to a table with the uncompressed value.}
    \label{f:simplified_tree}
\end{figure}

A significant issue when decoding a Huffman encoded sequence of bits is that it requires optimized lookup tables or complex hardware if high throughput is needed~\cite{phuffman}. In this regard, we proposed to use a simplified Huffman tree where we limit the tree size to a small number of nodes (i.e., four), as shown in Figure~\ref{f:simplified_tree}. In this tree,  several encoded $bit~sequences$  are assigned to each node. Note that we could employ a table to store the encoded $bit~sequences$ belonging to a given node. Hence, during the decoding process, we use the first bits of the encoded $bit~sequence$ to find out to which table (node) it belongs and then employ the remaining bits to address the table and get the uncompressed $bit~sequence$. This simplified version provides a good trade-off between simplicity and compression rate, as shown in Section~\ref{s:results}.

\subsection{Removing less frequent $bit~sequences$}\label{s:removing_less_common}

We observe that in some cases, a rarely used $bit~sequence$ ($sa$) can be replaced by one employed more frequently ($sb$) without negatively affecting the model accuracy. For this, we constrain the hamming distance between $sa$ and $sb$ to 1 (i.e., only a one-bit difference ) to keep the error introduced during the evaluation low. Hence, we can further improve the compression ratio by increasing the frequency of use of the most common $bit~sequences$ and removing some of the less common ones. To exploit this observation, we employ the following algorithm.

First, for all the 3x3 binary kernels in a given basic block, we create a set $st$ containing the $M$ most commonly used $bit~sequences$. Then, we create a second set $su$ containing the $N$ less commonly employed. After this, we try to replace the $bit~sequences$ in $su$ with $bit~sequences$ from $st$. This process is done by first taking a $bit~sequence$ ($sa$) from $su$. Then comparing its hamming distance with the hamming distance of the $bit~sequences$ in $st$. When a $bit~sequence$ ($sb$) is found, such that the hamming distance between $sa$ and $sb$ is one, we replace $sa$ by $sb$. Note that if many $bit~sequences$ match this criterion, we employ the $bit~sequence$ with the highest frequency to replace $sa$. Finally, we repeat this process for all the $bit~sequences$ in $su$. We empirically searched for some combinations of $M$ and $N$; while this search is not exhaustive, we found some combinations that significantly improve the compression ratio, shown in Section~\ref{s:results}.

\section{Hardware and Software Implementation}

This section describes our proposal to improve BNN performance using compressed kernels. First, we present an overview of our proposal. Next, we detail the software baseline used to evaluate the model. Then, we motivate the need for some hardware optimizations. Finally, we discuss the hardware implementation of our scheme.

\begin{figure}[t!]
    \centering
    \includegraphics[width=3.375in]{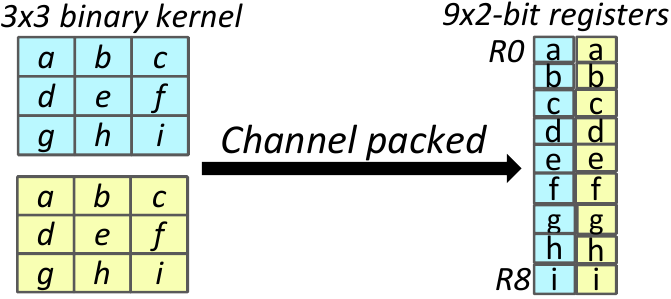}
    	\vspace*{-2mm}
    \caption{Channel packing a 2-channel 3x3 binary kernel. In this case, a 2-bit register is used to store a bit from each channel. }
    \label{f:packing_ex}
\end{figure}

\subsection{Overview}\label{s:overview}
One of the goals of this work is to employ the compressed 3x3 binary kernels during the evaluation of the 
BNN model. Our overall proposal work as follows. First, we compute the frequency of use for each $bit~sequence$ in all the 3x3 binary kernels. Then, based on those frequencies, we create a Huffman Tree and assign an encoding to each of the $bit~sequences$. Note that the previous steps are done offline and only once. 
Next, when evaluating a 3x3 kernel, we store the Huffman tree corresponding to that kernel in memory. After this, we start fetching the compressed $bit~sequences$ and inputs from memory. Then, the $bit~sequences$ are decompressed, and the kernel
evaluation is performed. This process is repeated for all $bit~sequences$ in the kernel. Nonetheless, compressing the kernels decreases the storage requirements; it adds an overhead to decode the $bit~sequences$ resulting in a slowdown when implemented in software. However, the overhead can be mitigated with additional hardware support. We detail these issues in the following sections.  

\begin{figure}[t!]
    \centering
    \includegraphics[width=3.375in]{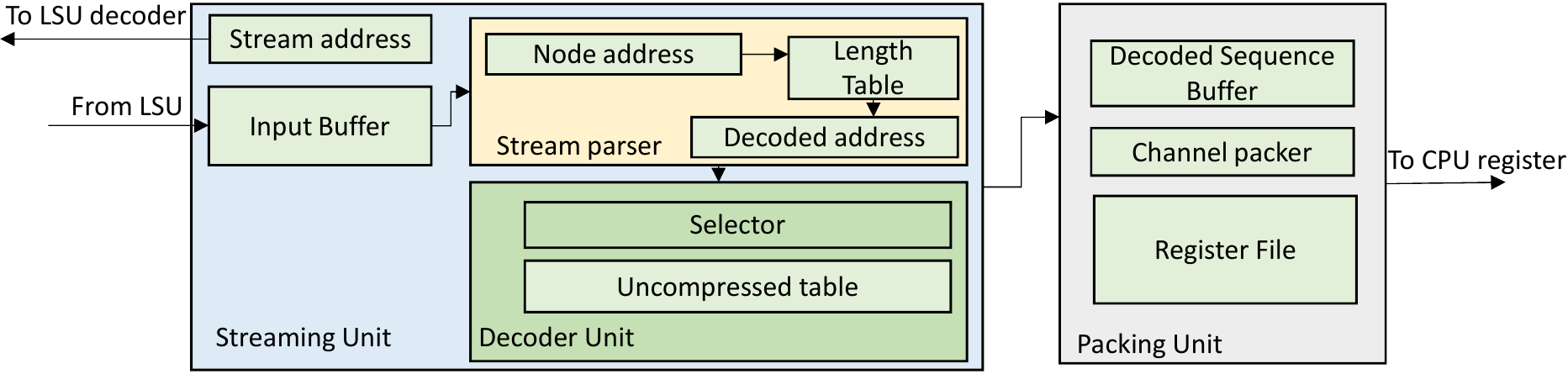}
    	\vspace*{-2mm}
    \caption{Decoding Unit}
    \label{f:decoding_unit}
\end{figure}

\subsection{Software Baseline}\label{s:software_baseline}

When evaluating BNNs in modern CPUs, one of the most critical components is laying out the kernels in memory such that the number of bits loaded into the internal register of the CPU~\cite{zhang2019dabnn,trusov2022fast} is maximized. In this regard, one of the most efficient methods is to pack bits from different channels together so that the number of bits packed together fills an entire register when loaded from memory~\cite{zhang2019dabnn}. For simplicity, we refer to this method as \textit{channel packing}.       

In \textit{channel packing}, the group of bits packed from each channel must belong to the same position (i.e., 0,0), as shown in Figure~\ref{f:packing_ex}. Also, when the number of channels is larger than the CPU register's size, the channels are split into several sub-groups such that they can be loaded independently into the CPU registers.
Note that \textit{channel packing} works best when the number of channels is divisible by the size of CPU registers; otherwise, padding is needed. Padding BNN kernels is challenging since the zero value represents a -1. Subsequently, when padding is done, the kernel computation must account for the extra -1. In this work, we employ \textit{channel packing} to store the uncompressed 3x3 binary kernels, and it is done offline, as in~\cite{zhang2019dabnn}. Also, since the number of channels for all the kernels is a power of two, we do not employ padding.

Regarding the layout in memory of the compressed kernels, note that each of the encoded $bit~sequence$ has a variable length, and thus we cannot perform channel packing offline. In this case, we store them consecutively in memory as a sequence of encoded words. As a result, during the kernel evaluation, once an encoded $bit~sequence$ is decoded, we need to pack it into the CPU registers. Note that the size of a decoded $bit~sequence$ is 9 bits, whereas the typical size of a CPU register is 32 bits or more; hence if packing is not done, the CPU registers are underutilized.

We implemented our compression scheme in software to evaluate its feasibility. Then, we compare the performance of this implementation against a baseline using uncompressed kernels. For the compression scheme, we implemented the simplified Huffman Tree described in Section~\ref{s:kernel_compression}. In this case, we limit the size of each tree to four nodes and use four tables to store the decoded $bit~sequence$. Then, we employ the first bits of each encoded $bit~sequence$ to indicate the table where it belongs. Moreover, we use the remaining bits in the uncompressed $bit~sequence$ to get the original from its corresponding table.
After comparing these two implementations, the compressed kernel implementation is 1.47x slower than the baseline.

\subsection{Hardware Support}\label{s:hardware_compression}

The main reason for the slowdown of the software implementation of our scheme is the overhead of decoding and packing the $bit~sequences$ at runtime. Hence, to mitigate this issue, we propose performing it in hardware. For this, we add a \textit{decoding unit} to the load-store unit (LSU) of the CPU. Furthermore, we add two new instructions to manage this unit. The \textit{decoding unit}, shown in  
Figure~\ref{f:decoding_unit}, is tailored to streaming and decoding the encoded \textit{bit sequences}. As can be seen, it is composed of a \textit{streaming unit} and a \textit{packing unit}.

The \textit{streaming unit} is employed to load a stream of encoded \textit{bit sequences} from the main memory and decompress them. It works as follows, first, it uses the \textit{stream address register} to store the base address of the \textit{bit sequences} that will be fetched. Then, a request to load $T$ bytes from the address in the \textit{stream address register} is sent to the LSU decoder. Once the $T$ bytes are fetched from memory, they are stored in the \textit{input buffer}. After this, the decoding process starts by sending $m$-bits of data from the \textit{input buffer} to the \textit{stream parser}. Note that, we also send a new request to fetch more bytes while doing the decoding. In the \textit{stream parser}, from this $m$-bits the first $n$ bits are used to create the \textit{node address} which identifies the node of the Huffman tree belonging to the current \textit{bit sequence}. Note that since the length of the encoded \textit{bit sequence} is variable, after finding the node, we need to find its length. It is done by reading it from the \textit{length table}, which is addressed with the \textit{node address}. Then, with the length information, we can get the remaining bits of the encoded \textit{bit sequence} and store them in the \textit{decoded address register}. Next, in the \textit{decoder unit}, the uncompressed \textit{bit sequences} are stored in a scratchpad memory (\textit{uncompressed table}), which is partitioned into multiple banks. Then, we employ the \textit{node address} and the value in the \textit{decoded address register} to address the \textit{uncompressed table} and get the uncompressed \textit{bit sequence}. Finally, the uncompressed sequence is passed to the $packing~unit$. We repeat the previous process until the whole stream of bits in the \textit{input buffer} have been processed.

The $packing~unit$ is in charge of \textit{channel packing} each one of the \textit{bit sequences} received from the $streaming~unit$ into the \textit{packing registers}. In this unit, after receiving a \textit{bit sequence}, it is distributed into $k$ register (i.e., 9 ) each of $R$ bits (i.e., 128 bits). Note that $R$ \textit{bit sequences} are channel packed sequentially into \textit{k} registers.
Also, once $R$ \textit{bit sequences} are packed, a new set of $k$ register is used to store the incoming \textit{bit sequences}.

To use the \textit{decoding unit}, the programmer first must configure it. For this, we add the instruction \textit{lddu} (i.e., load decoder unit configuration). This instruction uses a pointer to a  configuration structure to load its values in the \textit{decoding unit}. Table~\ref{t:configuration_structure} shows the main configuration values for the \textit{decoding unit}  which are stored in the main memory. Once the configuration values are loaded, the \textit{decoding unit} is reset and it starts decoding the \textit{bit sequences}.    

For reading the packed and uncompressed \textit{bit sequences} from the \textit{decoding unit}, we add the instruction \textit{ldps} (i.e., load packed bit sequence). To use this instruction, the programmer must specify the destination register to store the \textit{bit sequence} being read. Note that this instruction will read the oldest \textit{bit sequence} that was decoded in the \textit{decoding unit}. Hence, the programmer is responsible for setting this unit before using \textit{ldps}.      

To summarize, the \textit{decoding unit} works as follows. First, a \textit{lddu} instruction is executed to start its configuration and the decoding process. Then, in the background, the \textit{decoding unit} fetches and decodes the compressed \textit{bit sequences}. Later, when evaluating a 3x3 convolution, the programmer uses the \textit{lddu} instruction to load the uncompressed and channel-packed \textit{bit sequences} into the CPU registers. This process is repeated during the computation of a kernel until all the \textit{bit sequences} are evaluated.

\begin{table}[t!]
	\caption{Configuration Structure.}
	\label{t:configuration_structure}
	\centering
	\begin{tabular}{cc}
		\hline
		\multicolumn{2}{c}{Number of bit sequences }\\
		\multicolumn{2}{c}{\cellcolor[gray]{0.9}\small{Compressed sequences pointer}}\\
			\multicolumn{2}{c}{\small Compressed sequences length }\\
			\multicolumn{2}{c}{\cellcolor[gray]{0.9}\small Huffman tree nodes}\\
	
	\end{tabular}
\end{table}
	\vspace*{-2mm}

\section{Evaluation Methodology}\label{s:methodology}
We employ the Pytorch implementation of ReActNet and the ImageNet dataset as our baseline to evaluate the accuracy. Also, we use Python to compress and analyze the frequency of use for each bit sequence. For the performance evaluation, we use the daBnn~\cite{zhang2019dabnn} C++ framework to implement the BNN model. DaBnn employs \textit{channel packing} for the kernels and inputs. Moreover, it provides several C++ functions to compute the most common layers found on BNNs. This library is highly optimized for the ARMv8 architecture and uses the NEON extensions.

As a hardware baseline, we use an ARM A53 CPU. We use Gem5 to implement and evaluate our hardware optimizations. To assess the software baseline with the proposed hardware extensions, we rewrote the assembly functions used by daBnn, such that we exploit them when computing the binary kernels. Also, to estimate the latencies of our hardware structures, we implement them in Verilog and use the Synopsys Design Compiler for synthesis. We modeled a 4GB DDR4 DRAM for the main memory. The main configuration parameters for the CPU and the \textit{decoding unit} are shown in Table~\ref{t:hardware_params}.

\begin{table}[t!]
	\caption{Configuration Parameters.}
	\label{t:hardware_params}
	\centering
	\begin{tabular}{cc}
		\hline
		\cellcolor[gray]{0.9}\small\textbf{Parameter}&\cellcolor[gray]{0.9}\small\textbf{Value}\\
		\small CPU&\small ARM A53\\
		\cellcolor[gray]{0.9}\small Main Memory&\cellcolor[gray]{0.9}\small 4 GB DDR4\\
		\small Vector Registers&\small 32 (128 bits)\\
		\cellcolor[gray]{0.9}\small Frequency&\cellcolor[gray]{0.9}\small 1 GHz\\
		\small CPU L1 Cache&\small 32KB\\
		\cellcolor[gray]{0.9}\small CPU L2 Cache&\cellcolor[gray]{0.9}\small 256 KB\\
		\hline
		\multicolumn{2}{c}{\cellcolor[gray]{0.9}\textbf{Decoding Unit }}\\
		
		\small Max number of Nodes&\small 4\\
		\cellcolor[gray]{0.9}\small Uncompressed table &\cellcolor[gray]{0.9}\small 1 KB\\
		\small Register file&\small 256 bytes\\
		\cellcolor[gray]{0.9}\small Input Buffer&\cellcolor[gray]{0.9}\small 256 bytes \\
		\hline
		
	\end{tabular}
\end{table}

\section{Results}\label{s:results}

This section presents the evaluation of the proposed scheme implemented on top of an ARM CPU. We limit the Huffman tree to four nodes. Also, we restrict the number of \textit{bit sequences} stored on each node to 32, 64, 64, and 256. Hence, we use 6 bits to store the 32 most common sequences. Similarly, we use 8,9 and 12 bits for the other nodes, respectively.

Table~\ref{t:compression_ratio_by_layer} shows the compression ratio by layer. In this table, the Encoding column refers to encoding the 3x3 kernels without removing the less frequent $bit~sequences$, whereas the 256 most uncommon are removed for the Clustering column. For the Encoding, the compression ratio is between 1.18x and 1.25. The improvements came from employing only 6 bits to store the 32 most common \textit{bit sequences}, which have an average frequency of use of 46\%. Moreover, the rest of the \textit{bit sequences} are stored using 8, 9, and 12 bits with a frequency of use of 24\%, 23\%, and 5\%, respectively. After removing some of the least used \textit{bit sequences}, the compression ratio is improved by 1.32x on average. These improvements are due to increasing the frequency of the top 32 \textit{bit sequences} to 65\% on average. In this case, the frequency of use of the stored sequences using 12 bits is 0.6\%, whereas, for 8 and 9, it is 25\% and 8\%, respectively. Finally, the overall BNN model is compressed by 1.2x.

Regarding the performance, we compare our compression scheme with hardware support with the software implementation of ReAcNet. In this case, we obtain a speedup of 1.35x. The primary source for this improvement is the reduced latency for loading the 3x3 kernels. Note that since the \textit{decoding unit} fetches some of the \textit{bit sequences} while decoding and evaluating others, the load's latency overlaps with the computations.

\begin{table}[t!]
	\caption{ Compression ratio of $bit~sequences$ for the 3x3 kernel in each basic block.}
	\label{t:compression_ratio_by_layer}
	\centering
	\begin{tabular}{cccc}
		\cellcolor[gray]{0.9}\small\textbf{Layer}&\cellcolor[gray]{0.9}\small\textbf{Encoding} &\cellcolor[gray]{0.9}\small\textbf{Clustering}\\
		\small Block 1 &\small 1.18&\small 1.30\\
		\cellcolor[gray]{0.9}\small Block 2 &\cellcolor[gray]{0.9}\small 1.22 &\cellcolor[gray]{0.9}\small 1.30\\
		\small Block 3 &\small 1.21&\small1.31\\
		\cellcolor[gray]{0.9}\small Block 4 &\cellcolor[gray]{0.9}\small 1.21&\cellcolor[gray]{0.9}\small 1.32\\
			\small Block 5 &\small 1.19& \small 1.30\\
		\cellcolor[gray]{0.9}\small Block 6 &\cellcolor[gray]{0.9}\small 1.20 &\cellcolor[gray]{0.9}\small 1.33 \\
		\small Block 7 &\small 1.18&\small 1.33\\
		\cellcolor[gray]{0.9}\small Block 8 &\cellcolor[gray]{0.9}\small 1.20&\cellcolor[gray]{0.9}\small 1.32\\
			\small Block 9 &\small 1.20& \small 1.31\\	
		\cellcolor[gray]{0.9}\small Block 10 &\cellcolor[gray]{0.9}\small 1.18 &\cellcolor[gray]{0.9}\small 1.32 \\
		\small Block 11  &\small 1.19&\small 1.33\\
		\cellcolor[gray]{0.9}\small Block 12 &\cellcolor[gray]{0.9}\small 1.25&\cellcolor[gray]{0.9}\small 1.36\\
		\small Block 13  &\small 1.22&\small 1.35    
				
	\end{tabular}
\end{table}

\section{Related Work}
Hardware acceleration for BNNs has been an active area of research in recent years.
In this regards, there have been several works targeting CPUs~\cite{trusov2022fast}, FPGAs~\cite{zhang2021fracbnn,du2022bnn,fiscaletti}, and ASICs~\cite{hosseini2021binary,cho2021reconfigurable}. FracBnn~\cite{zhang2021fracbnn} proposes a FPGA accelerator for ReActNet. In addition, it improves performance by employing up to two bits for activation and increasing the level of sparsity in the kernels. Also, they binarize the first layer of the model, wheres it is quantized in this work. The work in~\cite{du2022bnn} proposes an FPGA accelerator that combines the xnor and accumulation to improve performance. BNNsplit~\cite{fiscaletti} is another accelerator that divides the evaluation of the BNN model across multiple FPGAs. Our proposal is different from these works, since we focus on compressing the kernels and we target mobile CPUs.

Hosseini et al.~\cite{hosseini2021binary} propose a programmable many-core accelerator targeting the primary operations found in BNN networks. The work in~\cite{cho2021reconfigurable} is another BNN accelerator that analyzes each layer and, based on the given parameters, reconfigures the hardware to increase parallelism. Both works target ASICs, whereas our work focuses on mobile CPUs.

The work in~\cite{trusov2022fast} proposes a new blocking algorithm to partition the binary kernels using a small block size such that the utilization of CPU vector registers is more efficient. In Deep Compression~\cite{han2015deep} Huffman encoding is applied to full-precision networks after pruning them. This scheme is proposed for GPUs, and the optimizations are in software. Our proposal is different since it is a hybrid software/hardware approach.

\section{Conclusion}
This work presents a novel compressing scheme for binary neural networks. We show that 
in the binary kernels, some \textit{bit sequences} occur more frequently than others. Aiming to exploit this observation, we encode each sequence with a simplified Huffman Tree. We propose a simple hardware extension to conventional CPUs to support this scheme, since implementing it on software degrades performance. After applying our compression scheme, the binary kernels are compressed by 1.32x on average, whereas the BNN model is compressed by 1.2x. Also, compared to the software baseline, our proposal achieves a speedup of 1.35x.

\bibliographystyle{IEEEtran}

\bibliography{references}

\begin{thebibliography}{10}
\providecommand{\url}[1]{#1}
\csname url@samestyle\endcsname
\providecommand{\newblock}{\relax}
\providecommand{\bibinfo}[2]{#2}
\providecommand{\BIBentrySTDinterwordspacing}{\spaceskip=0pt\relax}
\providecommand{\BIBentryALTinterwordstretchfactor}{4}
\providecommand{\BIBentryALTinterwordspacing}{\spaceskip=\fontdimen2\font plus
\BIBentryALTinterwordstretchfactor\fontdimen3\font minus
  \fontdimen4\font\relax}
\providecommand{\BIBforeignlanguage}[2]{{%
\expandafter\ifx\csname l@#1\endcsname\relax
\typeout{** WARNING: IEEEtran.bst: No hyphenation pattern has been}%
\typeout{** loaded for the language `#1'. Using the pattern for}%
\typeout{** the default language instead.}%
\else
\language=\csname l@#1\endcsname
\fi
#2}}
\providecommand{\BIBdecl}{\relax}
\BIBdecl

\bibitem{reactnet}
Z.~Liu, Z.~Shen, M.~Savvides, and K.-T. Cheng, ``Reactnet: Towards precise
  binary neural network with generalized activation functions,'' in
  \emph{European Conference on Computer Vision (ECCV)}, 2020.

\bibitem{zhang2021fracbnn}
Y.~Zhang, J.~Pan, X.~Liu, H.~Chen, D.~Chen, and Z.~Zhang, ``Fracbnn: Accurate
  and fpga-efficient binary neural networks with fractional activations,'' in
  \emph{The 2021 ACM/SIGDA International Symposium on Field-Programmable Gate
  Arrays}, 2021, pp. 171--182.

\bibitem{mobilenetv2}
M.~Sandler, A.~Howard, M.~Zhu, A.~Zhmoginov, and L.-C. Chen, ``Mobilenetv2:
  Inverted residuals and linear bottlenecks,'' in \emph{Proceedings of the IEEE
  conference on computer vision and pattern recognition}, 2018, pp. 4510--4520.

\bibitem{phuffman}
Z.~Aspar, Z.~Mohd~Yusof, and I.~Suleiman, ``Parallel huffman decoder with an
  optimized look up table option on fpga,'' in \emph{2000 TENCON Proceedings.
  Intelligent Systems and Technologies for the New Millennium (Cat.
  No.00CH37119)}, vol.~1, 2000, pp. 73--76 vol.1.

\bibitem{zhang2019dabnn}
J.~Zhang, Y.~Pan, T.~Yao, H.~Zhao, and T.~Mei, ``dabnn: A super fast inference
  framework for binary neural networks on arm devices,'' in \emph{Proceedings
  of the 27th ACM international conference on multimedia}, 2019, pp.
  2272--2275.

\bibitem{trusov2022fast}
A.~Trusov, E.~Limonova, D.~Nikolaev, and V.~V. Arlazarov, ``Fast matrix
  multiplication for binary and ternary cnns on arm cpu,'' \emph{arXiv preprint
  arXiv:2205.09120}, 2022.

\bibitem{du2022bnn}
G.~Du, B.~Chen, Z.~Li, Z.~Tu, J.~Zhou, S.~Wang, Q.~Zhao, Y.~Yin, and X.~Wang,
  ``A bnn accelerator based on edge-skip-calculation strategy and consolidation
  compressed tree,'' \emph{ACM Transactions on Reconfigurable Technology and
  Systems (TRETS)}, vol.~15, no.~3, pp. 1--20, 2022.

\bibitem{fiscaletti}
G.~Fiscaletti, M.~Speziali, L.~Stornaiuolo, M.~D. Santambrogio, and D.~Sciuto,
  ``Bnnsplit: Binarized neural networks for embedded distributed fpga-based
  computing systems,'' in \emph{Proceedings of the 23rd Conference on Design,
  Automation and Test in Europe}, ser. DATE '20.\hskip 1em plus 0.5em minus
  0.4em\relax San Jose, CA, USA: EDA Consortium, 2020, p. 975–978.

\bibitem{hosseini2021binary}
M.~Hosseini and T.~Mohsenin, ``Binary precision neural network manycore
  accelerator,'' \emph{ACM Journal on Emerging Technologies in Computing
  Systems (JETC)}, vol.~17, no.~2, pp. 1--27, 2021.

\bibitem{cho2021reconfigurable}
J.~Cho, Y.~Jung, S.~Lee, and Y.~Jung, ``Reconfigurable binary neural network
  accelerator with adaptive parallelism scheme,'' \emph{Electronics}, vol.~10,
  no.~3, p. 230, 2021.

\bibitem{han2015deep}
S.~Han, H.~Mao, and W.~J. Dally, ``Deep compression: Compressing deep neural
  networks with pruning, trained quantization and huffman coding,'' \emph{arXiv
  preprint arXiv:1510.00149}, 2015.

\end{thebibliography}

\end{document}